\documentclass[twocolumn,showpacs,preprintnumbers,amsmath,
amssymb]{revtex4}
\usepackage{graphicx}
\usepackage{dcolumn}
\usepackage{bm}
\usepackage{tensor} 
\begin{document}

\title{Variational Principle for the Optical Phase.}
 
\author{A.Yu.Okulov}
\email{alexey.okulov@gmail.com}
\homepage{https://sites.google.com/view/okulovalexey11/bio}
\affiliation{Russian Academy of Sciences, 119991,  Moscow, 
Russian Federation.}

\date{\today} 

\begin{abstract}
{The problem of laser beam concentration in a focal 
spot via wavefront variations is formulated 
as a maximization of the $beam$ $propagation$ $functional$ defined as the 
light power passing through aperture 
of an arbitrary shape located in the far field. 
Variational principle provides the necessary and sufficient conditions  
for at least the $local$ $maximum$ of the $beam$ $propagation$ $functional$. 
The wavefront shape is obtained as an exact solution of nonlinear integral equation. }
\end{abstract} 

\maketitle

\vspace{1cm}

\section {Introduction}

Interdisciplinary methods introduced by Haken \cite{Haken:1975} 
stem from analogy \cite {Haken:1975pla} between low-dimensional 
hydrodynamical Lorenz flow \cite {Lorenz:1963} and laser 
high gain instabilities \cite {Oraevsky:1964, Oraevsky:1981}. 
The unified view on spatiotemporal 
structures formed in nonequilibrium systems, appearance of chaos, solitons and 
pattern formation in the presence of fluctuations \cite{Cross:1993} 
resulted in significant technological achievements in recent decades. 
Starting from the 
early theory of soliton mode locking in a general model of laser with the fast 
saturable absorber \cite{Haus:1975} the subsequent efforts \cite{Krausz:1996} 
stimulated successful experimental realizations. One of the most visible 
manifestations of this interdisciplinary approach proved to be 
the massive implementation of femtosecond soliton lasers 
\cite{Keller:1997} based upon diode-pumped thin disk laser 
concept \cite {Okulov:1990,Giesen:1994}  
known earlier also as the active mirror laser \cite {Basov:1966,Bogdankevich:1973}. 
The other promising experimental field emerged due to the successful development 
of the femtosecond chirped pulse fiber laser arrays \cite{Mourou:2013} 
where the accurate phase locking of a large mode area (LMA) fiber amplifiers  
provides an impressive coherent beam combination with a high repetition rate. 
In both above cases the substantial $phase$ $fluctuations$ induced by 
imperfectness of equipment and random noise were put under reliable control.
The optical phase control is in the heart 
of laser beam combination. 

Since the early years of the laser science a  
beam combination became a subject of the vital research interest 
\cite {Letokhov:1965,Lamb:1972}. 
Phase-locked arrays of gas, semiconductor \cite {Napartovich:1990},
fiber \cite {Mourou:2013} 
and diode-pumped thin-disk solid-state lasers \cite {Okulov:1990} 
had been studied from the point of view of a careful adjustment 
of the flattened wavefronts and a path-differences lag with 
$\lambda/(10 \div 100)$ accuracy. The shape of the 
output wavefront of the phase-locked arrays is crucial for a free-space 
laser transmitters and many other applications including 
materials processing. The currently available direct method  
of a laser array phase-locking is 
Shack-Hartman technique developed originally for 
correction of the profile of mirrors of a large telescopes. Nowadays the  
Shack-Hartman technique have shown to be highly effective tool 
for the adjustment of a relative phases of a dozens of amplifying channels 
in phase-locked fiber arrays \cite{Mourou:2013}. 
In fact the Shack-Hartman technique is a variational method  
of maximization the intensity in the far field by means of 
careful tuning of the optical paths in a near field . 
This technique requires control of the large number 
of phase delays by a high performance computers \cite{Okulov:2014}. 
The typical experimental configuration 
used for phase-locking of the chirped pulses being amplified 
by a dozens of the LMA 
fiber amplifiers comprises ethalon master oscillator with transform-limited  
few-period femtosecond output ($\tau_{osc} \sim 30 \div 50 fs$), stretcher,  beam splitter tree, network of 
fiber amplifiers, the phase-piston correctors, beam combination network for 
amplified pulses with a relatively long duration ($\tau_{amp} \sim 30 \div 900 ps$) and the 
pulse compressor at final stage.

Current communication is devoted 
to optimization of the output phase profile of a phase-locked 
laser sets \cite{Basov:1980} and large area mirrors within framework of the 
properly formulated variational 
principle which results in exact solution for the optimal 
wavefront shape. 

\section {Variational principles for the trajectories of particles,
rays and pattern formation.}

Variational principles play an important role in formulation 
of equations of motions in different areas beginning from 
classical to quantum mechanics and beyond. The Hamiltonian principle 
of the least action $S$ provides equations of motion 
 for relativistic charged 
particle in electromagnetic field \cite {Landau:1982, Greiner:1996} (fig.1):
\begin{eqnarray} 
\label {Least Action} 
\delta S = \delta \int_{t_1}^{{t_2}} 
{L(\vec r, \dot{\vec{r}},t)}dt = 
& \nonumber \\  
\delta \int_{t_1}^{t_2} 
[{-mc^2}\sqrt{1-\frac {v^2}{c^2}}+{\frac{e}{c}}
{\vec A}(\vec r,t)\cdot \dot{\vec{r}}-e\phi(\vec r,t)]{\:}dt=0,
\end{eqnarray}
where $m$, $e$ are the mass and electric charge of a particle, 
$c$ is the speed of 
light, $\vec A$ is vector potential, $\phi$ is scalar 
potential, $ v=|\vec {\dot{r}}|$, 
Langrangian $L$ reduces in nonrelativistic limit $v^2<<c^2$ 
to the difference of kinetic and potential energy $L=T_{kin}-U$ 
inherent to classical mechanics.

Maupertuis variational principle for the nonrelativistic particle of 
a mass $m$ and electric charge $m$ 
in constant electromagnetic field with vector 
potential ${\vec A}(\vec r)$ \cite {Landau:1982} is known as: 

\begin{equation}
\label {Maupertuis} 
\delta \int {{\vec P}\cdot {\vec {dr}}}=0,
\delta \int {{\vec p}\cdot {\vec {dl}}}+{\frac {e}{c}}
{{\vec A}(\vec r)}\cdot {d \vec r}=0,
\end{equation}
 
where ${\vec {dl}}$ is an element of the particle trajectory.
\begin{figure} 
\center{ \includegraphics[width=9.cm]{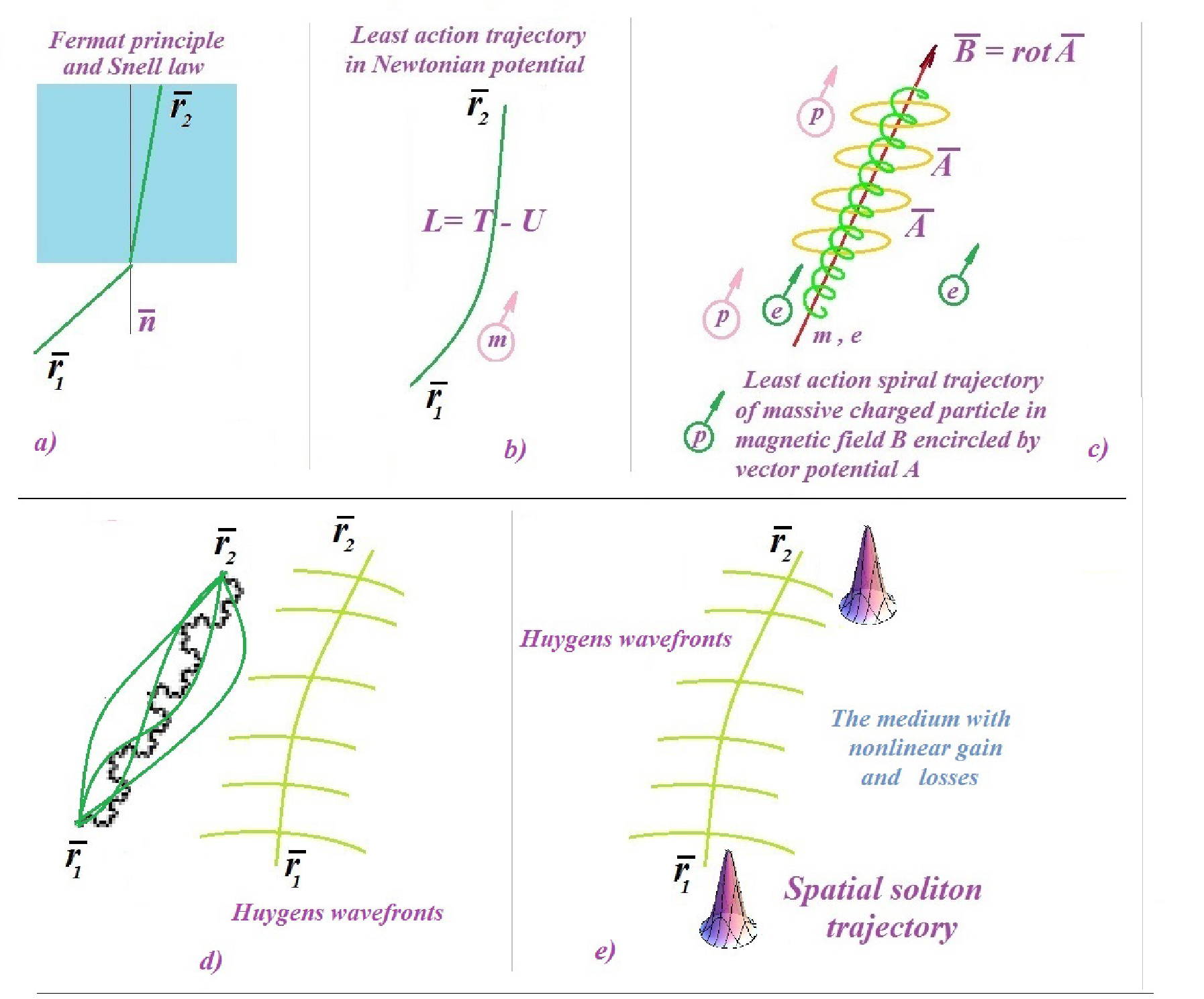}}
\caption{ (Color online) Different variational principles for 
different situations and geometries. a) Fermat principle 
for refraction of rays at the boundary. b) Hamilton 
principle for massive particle $m$ in Newtonian 
potential and Lagrangian $L=T-U$. 
c) Variational principle for the relativistic massive $m$ charged $\pm e$ particles 
moving from point $\vec r_1$ to point $\vec r_2$ in magnetic field $\vec B=rot \vec A$ 
encircled by vector potential $\vec A$. 
d) Feynman path integral variational principle for particle 
moving from point $\vec r_1$ to point $\vec r_2$ whose classical 
trajectory appears as a result of constructive interference of adjacent paths (green).
The origin of classical trajectory may be considered as a result of constructive 
interference of partial $linear$ waves 
emitted by Huygens-Frenel wavefronts. 
The fractal curve connecting points $\vec r_1$ and $\vec r_2$ is a trajectory 
of particle that appears under attempts of continuous measurements. The evaluation  
of Hausdorf fractal dimension $D_H$ in this case gives $D_H=2$ \cite{Abbot:1981} .
e) Variational principle for the spatial soliton \cite{Malomed:2002}.  
The particle-like localized excitation \cite {Okulov:1988} 
moves in a nonlinear medium 
with gain and losses from point $\vec r_1$ to point $\vec r_2$. The 
trajectory of soliton is a result of constructive interference of partial $nonlinear$ waves 
emitted by Huygens-Frenel wavefronts.} 
\label{fig.1} 
\end{figure}

When electromagnetic potentials are absent the Maupertuis principle 
reduces to the Fermat principle for the least time of 
propagation of light and particles between 
points $\vec r_1$ and $\vec r_2$ and reads as follows \cite {Landau:1982}(fig.1): 
 
\begin{equation}
\label {Least Time Fermat } 
\delta \psi = \delta  \int_{{\vec r_1}}^{{\vec r_2}} 
({\vec k} \cdot {\vec {dl}}) = 0, {\:}{\:} \vec k=\frac {\partial \psi}{\partial \vec r} 
= \vec \nabla \psi, {\:}{\:}{\:}{\:}
\end{equation}

where $\psi $ is eikonal, $\vec k= grad {\:} \psi$ is wavevector.
 The principle of the least action $S$ provides the clear picture of 
the quantum-classical 
correspondence in the Feynman formulation of quantum mechanics where 
classical trajectories arise 
as a result of the constructive interference of probability amplitudes 
for the continuum set of the trajectories 
linking two points $\vec r_1, \vec r_2$ 
reached by the particle at the moments  $t_1, t_2$ \cite {Feynman:1965}(fig.1):
  
\begin{equation} 
\label {Feynman } 
\Psi (\vec r_1, \vec r_2)  = \int_{\vec r_1}^{\vec r_2} \lbrace
\exp{ \frac {i  }{\hbar}}\int_{t_1}^{t_2} {L(\vec r, \dot{\vec{r}},t)}dt 
\rbrace {\:}{\:}{\mathcal D} \vec r(t),
\end{equation}

where $\Psi (\vec r_1, \vec r_2)$ is Dirac probability amplitude 
for particle moving from the point $\vec r_1$ to the point $\vec r_2$, 
the measure ${\mathfrak D} \vec r(t)$ is an infinitesimal subset 
of the $all$ $possible$ 
$trajectories$ including the $nondifferentiable$ $trajectories$
(\cite {Feynman:1965}, Ch.7, fig.$7.1$) provided these trajectories 
connect the starting and final points  $\vec r_1, \vec r_2$. 
It had been shown that during continuous set of simultaneous 
measurements of particle position and velocity the resulting 
trajectory of quantum mechanical particle may be a fractal due to 
Heisenberg uncertainty relations and such a trajectory  
has a Hausdorf dimension $D_H$ equals two \cite{Abbot:1981}.
  
In the field of spontaneous pattern formation in nonequilibrium 
dissipative systems the variational principle had been introduced by 
Haken as a minimization of free-energy functional \cite{Haken:1975}. 
The similar variational principles were formulated in the field of laser 
dynamics in \cite{Akhmanov:1981,Lugiato:1991} and for the  
least entropy production by Prigogine \cite{Prigogine:1978}.

Variational principle in conventional form 
$\delta S = \delta \int L dz$ had been formulated 
for the nonlinear evolution partial differential equations alike 
nonlinear Shrodinger-Ginzburg-Landau equation\cite{Malomed:2002} with 
Maxwell-Bloch gain:
\begin{eqnarray}
\label {envelope for gain loss medium} 
\frac {\partial E(z, \vec {r},t)}{\partial z} + {\frac {1}{c}} \frac {\partial E(z, \vec {r},t)}{\partial t} + 
{\frac {i}{2k}}\Delta_{\bot} E(z, \vec {r},t)= 
& \nonumber \\ 
-\gamma E(z, \vec {r},t) + ik^2 n (z, \vec {r}) E(z, \vec {r},t)
+ \sigma N (z, \vec {r},t) E(z, \vec {r},t) ,{\:}{\:} 
& \nonumber \\ 
\frac {\partial N (z, \vec {r},t)}{\partial t} + \frac {N_0(z, \vec {r},t)}{T_1} = 
\sigma {N (z, \vec {r},t)}\cdot \epsilon_0 c | E(z, \vec {r},t)|^2, {\:}{\:}
& \nonumber \\ 
\sigma = \frac {\sigma_0} {1+({\Delta \omega} \cdot{T_2})^2}{\:}{\:}{\:}{\:}{\:}{\:}
\end{eqnarray}
 
where $z$ is variable along beam propagation trajectory , ${E(z, \vec {r},t),N (z, \vec {r},t)}$ are electric 
field and population inversion envelopes, $T_1$,$T_2$ are the longitudinal 
and transverse relaxation times \cite {Oraevsky:1964, Oraevsky:1981},
$\epsilon_0$ is the dielectric permeability of vacuum, $c$ is the speed of light, 
${\Delta \omega}$ is detuning of carrier frequency from gain linewidth, 
${N_0(z, \vec {r},t)},n (z, \vec {r}) $ are distributions of gain and index,
$\sigma$ is stimulated emission cross-section.  
This variational principle is applicable for description of 
 solitons in nonlinear optical 
media and Bose-Einstein condensates: 
 
\begin{equation}
\label {Least action malomed } 
S =  \int_{z_1}^{{z_2}} Ldz; {\:}{\:} L=\int \mathcal L
({\vec r} , {\tau}) d{\tau}d^2{{\vec r}},{\:}
\end{equation}
where $\mathcal L$ is Lagrangian density. This approach gives the correct subset 
of the ordinary differential equations for parameters of the  
 spatial soliton envelopes.
 
The equations of motion for parameters of trial envelope functions 
provide a satisfactorily agreement with numerical modeling 
and experiments with ultrashort pulses in fibers and 
localized excitations in ultracold atomic ensembles.

The spatial soliton motion in a gain-loss medium might be interpreted as 
$nonlinear $ Huygens-Fresnal principle where each current wavefront in 
a given moment of the evolutional variable $\tau=t - z\cdot c$  emits 
the continuum of partial $nonlinear$ Huygens-Fresnel waves whose constructive 
interference results in a localized self-similar wavetrain moving across medium (fig.1).  

\section {Beam propagation functional in the explicit form} 

For the purpose of achieving the $maximum$ laser $power$ $concentration$  
in the far field or in the focal plane of an ideal lens (fig.2)  
one may formulate the variational principle in the framework 
of the paraxial wave optics \cite {Born_Wolf:1972} and 
mathematically equivalent nonrelativistic wave mechanics (\cite {Feynman:1965}. 
This approach differs 
from previously found solution of achieving the $perfect$ drawing 
of $given$ line in the $near$ $field$ or inside the Fresnel diffraction 
zone where the exact match of laser 
beam intensity distribution and demanded line shape on a plane had been 
realized 
by means of the perfect arrangement of a $zeros$ of optical field 
$E(z, \vec {r})$ \cite {Volostnikov:1989} known as 
optical vortices \cite {Allen:1992,Gbur:2016}. 

It seems reasonable to seek 
a maximum of the optical flux passing through 
the aperture of an arbitrary shape $D(\vec r)$ in a far field. 
In a simplest situation of the scalar wave optics and homogeneous 
polarization distribution across the laser beam transverse 
section the optical flux transmitted through 
aperture placed in a far field $D(\vec r)$ is given by: 
\begin{equation}
\label {TFT} 
T [\psi (\vec r,z) ]  =   \int_{-\infty}^{\infty} D(\vec r) I (\vec r,z)  
d^2{\vec r}{\:};{\:}\frac {\delta T [\psi (\vec r,z)]}{\delta \psi}=0,
\end{equation}
where $I (\vec r,z \rightarrow +\infty)$ is intensity of laser 
radiation in the far field at 
$z=\infty$ or in the focal plane of perfect lens at $z=F$.   
The near field emission intensity at $z=0$ as $I (\vec r,z=0)$ 
has a complex 
electric field amplitude 
$E(z=0, \vec {r}')= A (\vec {r}')\exp [i \psi (\vec {r}')]$ with 
amplitude $A (\vec {r}')$ and phase $\psi (\vec {r}')$ distributions 
as standard boundary conditions on a plane $z=0$ in Cauchi problem. The 
diaphragm transmittance $D(\vec r)$ is dimensionless real function 
whose values are within $[0,1]$ range while 
$\psi (\vec r) $ is a phase mask of an arbitrary profile which 
may be a smooth function or a purely chaotic random phase 
plate \cite {Okulov:2014,Basov:1980} which may be modeled numerically 
as a multimode random process 
\cite {Okulov:1991}. We will seek for the 
$necessary$ and $sufficient$ conditions \cite{Greiner:1996} for the 
extremum of $target$ functional $T [\psi (\vec r,z)]$. Extremum is defined from  a 
$nesessary$ condition $\frac {\delta T [\psi]}{\delta \psi}=0$ on the $L_2$ 
(Lebesgue) functional space 
of piecewise continuous phase mask functions $\psi (\vec r,z)$ 
with the integrable square $|\psi (\vec r,z)|^2$ \cite {Vladimirov:1971}. 
The $sufficient$ condition for second variational 
derivative $\frac {\delta^2 T [\psi]}{{\delta \psi}^2}{\:} \lessgtr 0 $ 
defines whether a given functional has a maximum or a minimum on a 
particular near field phase function $\psi (\vec r,z=0)$ \cite{Greiner:1996}. The hardware 
implementation is electrooptically controlled adaptive phase mask (fig.3). 
The optimal phase mask $\psi (\vec r,z=0)$ ensures constractional interference 
in a far field or in the binary-tree 
beam combination schemes \cite {Basov:1980}.
 \begin{figure} 
\center{ \includegraphics[width=9 cm]{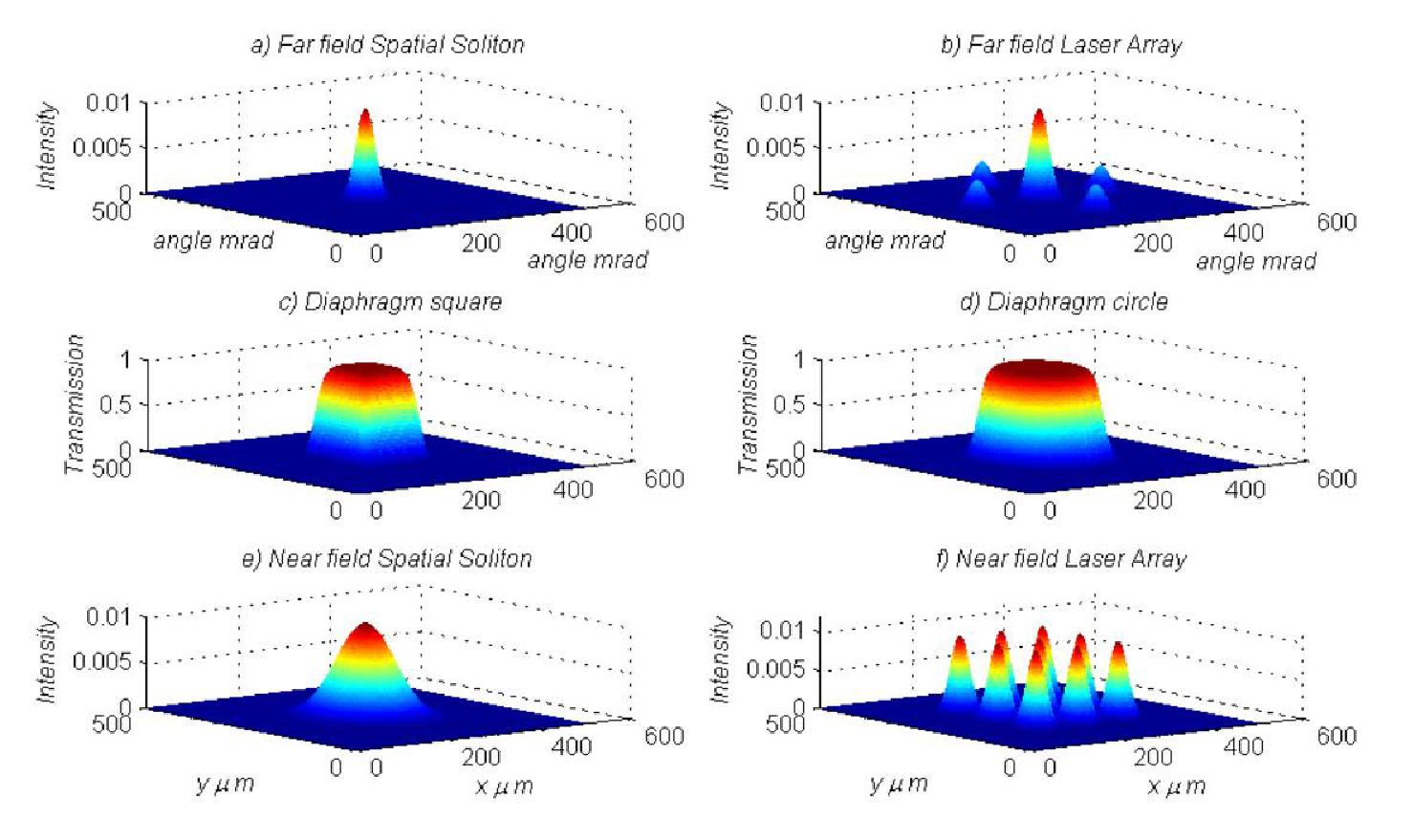}}
\caption{ (Color online) Composition of $target$ $functional$.
a) Far field of $spatial$ $soliton$ \cite {Okulov:2020} 
or a Gaussian fundamental cavity mode.
b) Far field of $laser$ $array$ (\ref{TFT from solutions }) \cite {Mourou:2013}.
c) $Rectangular$ diaphragm in far field $D({\vec r})$.
d) $Circular$ diaphragm in far field $D({r})$.
e) Near field of $spatial$ $soliton$ \cite {Okulov:1988} 
or a Gaussian fundamental cavity mode.
f) Near field of thin disk $laser$ $array$ \cite {Okulov:1990} or 
fiber laser coherent network \cite{Mourou:2013}.} 
\label{fig.2}
\end{figure}
\subsection {Cartesian coordinates}

The variational derivative over the 
set of phase distributions $\psi (\vec r)$ means the search 
of the extremum (stationary point) of above defined 
$target$ functional $T [\psi ]$ 
under condition of $fixed$ $ {\:}amplitude$ of 
light $A (\vec {r}')$ in the near field (fig.2). 
The most frequently used near field amplitude distributions are 
fundamental Gaussian beams, Hermite-Gaussian and 
Laguerre-Gaussian modes \cite {Allen:1992,Gbur:2016},
Bessel and Bessel-Gaussian beams, spatial solitons \cite{Malomed:2002} 
obtained as transverse mode-locking in lasers \cite{Okulov:2020}. 
The spatial  solitons  with $sech$ transverse profile \cite{Moloney:1983} have 
the same  $sech$ far field \cite{Okulov:2000}. 
The very interesting spatially periodic field distributions \cite{Okulov:2014}
were already realized in phase-locked laser arrays \cite{Mourou:2013}.
The $fixed$ distribution 
of light amplitude $A (\vec {r}')$ at $z=0$ means the conservation 
of the total light intensity for all $z$ before passage through aperture 
$D(\vec r)$ in far field 
under arbitrary variations 
$\delta \psi (\vec {r}')$ in the near field:
\begin{equation}
\label {norm conservation } 
\frac {\partial {[\int I (z, \vec r)} d^2 {\vec r}]}{\partial z}=0 ; {\:}{\:} 
T [\psi (\vec r,z)  ]  =  {\int}_{-\infty}^{\infty} D(\vec r) I (\vec r,z)  
d^2{\vec r} .
\end{equation}
 The exact solution of the paraxial wave equation 
for $E(z, \vec {r})$ :
\begin{equation}
\label {TFT 2} 
{ik}\frac {\partial E(z, \vec {r})}{\partial z} + 
\Delta_{\bot} E(z, \vec {r})+ k^2 n (z, \vec {r}) E(z, \vec {r})= 0 {\:}{\:} ,
\end{equation}
in free space ($n (z, \vec {r})=1$) for an arbitrary plane $z$ 
reads as follows:
\begin{equation}
\label {TFT 3} 
E(z, \vec {r})= \frac {ik}{2 \pi z} e^{ikz}  \int d^2 \vec {r}' 
 \exp \lbrace \frac {ik}{2z}|\vec {r}'-\vec r|^2 \rbrace E(z=0,\vec {r}') {\:}{\:} ,
\end{equation}
where $k=2\pi/\lambda$ is wavenumber.

Consequently the $target$ functional has the following form:
\begin{eqnarray}
\label {TFT from solutions } 
T [\psi (\vec r),z \rightarrow \infty  ]=T [\psi ]  =  \int_{-\infty}^{\infty} D(\vec r) I (\vec r,z)  
d^2{\vec r}=
& \nonumber \\
\frac {\epsilon_0 c{\:} k^2}{4 \pi^2 z^2}  \int_{-\infty}^{\infty} D(\vec r)   
d^2{\vec r} \int_{-\infty}^{\infty} \int_{-\infty}^{\infty}
E(z=0,\vec {r}')E^*(z=0,\vec {r}'')
& \nonumber \\ 
\exp \lbrace \frac {ik}{2z}|\vec {r}'-\vec r|^2 
-|\vec {r}''-\vec r|^2\rbrace
d^2{\vec r''}d^2{\vec r'} \ge 0{\:}{\:} , {\:}{\:} {\:}{\:} 
\end{eqnarray}
therefore $T [\psi (\vec r),z \rightarrow \infty  ] \equiv T [\psi ] $ 
is $positively$ $defined$ (next for brevity we use  $T[\psi]$ as well) .

In the far field 
where $\exp {\frac {ik}{2z}|}\vec {r}|^2 \cong 0$ 
the above integrals are simplified as follows:

 \begin{eqnarray}
\label {TFT from solutions2 } 
T[\psi]  =  \frac {\epsilon_0 c{\:} k^2}{4 \pi^2 z^2} 
 \int_{-\infty}^{\infty} D(\vec r)   
d^2{\vec r} \int_{-\infty}^{\infty} \int_{-\infty}^{\infty}
A(\vec {r}')A(\vec {r}'')
& \nonumber \\ 
\exp \lbrace i [\psi(\vec {r}')-\psi(\vec {r}'')] 
+\frac {ik}{2z}(\vec {r}'-\vec {r}'')\cdot \vec r
\rbrace 
d^2{\vec {r}''}d^2{\vec {r}'}{\:}{\:} . {\:}{\:} {\:}{\:} 
\end{eqnarray}

The integration over $\vec r {\:}$  in the far field provides the 
simplified form of the functional $T[\psi]$:

 \begin{eqnarray}
\label {TFT from solutions3 } 
T [\psi ]  =  \frac {\epsilon_0 c{\:} k^2}{4 \pi^2 z^2} 
\int_{-\infty}^{\infty} \int_{-\infty}^{\infty}
A(\vec {r}')A(\vec {r}'')D_{iFT}(\vec {r}'-\vec {r}'')
& \nonumber \\ 
\exp \lbrace i [\psi(\vec {r}')-\psi(\vec {r}'')] 
\rbrace 
d^2{\vec {r}''}d^2{\vec {r}'}{\:}{\:} , {\:}{\:} {\:}{\:} 
\end{eqnarray}
where kernel $D_{iFT}(\vec {r}'-\vec {r}'')$ is $inverse$ 
Fourier transform 
of $D(\vec r)$ or diaphragm located in a far field as it 
viewed from plane $z=0$.  

First of all it is noteworthy to stress the point that functional 
$T [\psi ]$ is positively defined for $all$ $functions$ 
$A(\vec {r})$ and $\psi(\vec {r})$ because 
it is integral in the far field of the light intensity $I(\vec r)$
being real function times aperture transmission $D(\vec r)$. 
For this reason we have for the real part of $T [\psi ]$ 
of power flux through $D(\vec r) $ unavoidably : 
 \begin{eqnarray}
\label {TFT from solutions 4} 
Re \lbrace {T [\psi ] \rbrace} = \frac {\epsilon_0 c{\:} k^2}{4 \pi^2 z^2} 
\int_{-\infty}^{\infty} \int_{-\infty}^{\infty}
A(\vec {r}')A(\vec {r}'')D_{iFT}(\vec {r}'-\vec {r}'')
& \nonumber \\ 
\cos\lbrace [\psi(\vec {r}')-\psi(\vec {r}'')] 
\rbrace 
d^2{\vec {r}''}d^2{\vec {r}'}{\:}{\:} . {\:}{\:} {\:}{\:} 
\end{eqnarray}
Evidently the imaginary part $Im (T[\psi ])$ of functional is zero because 
it is integral over far-field intensity times diaphragm transmission:
 \begin{eqnarray}
\label {TFT from solutions5 } 
Im (T[\psi ]) {\:}={\:}0 {\:}.{\:} {\:} {\:}{\:} {\:}   
\end{eqnarray}
Noteworthy for infinitely wide far field aperture when 
$D(\vec r)=1$ its Fourier image $D_{iFT}$ is 
exactly delta-function $D_{iFT}\sim \delta( \vec {r}'-\vec {r}'')$
and both identities are solely real or imaginary.
In order to get exact solution of this optimization problem let us 
consider the symmetrical initial field distributions 
$A(\vec {r}'),A(\vec {r}'')$ and symmetrical aperture 
$D(\vec r)$ placed at the beam axis.  

\subsection {Cylindrical coordinates}

In cylindrical coordinates 
$ (x,y,z) \cong (\vec r,z) \Rightarrow (r, \theta,z) $ 
the paraxial wave equation 
for $E(z, r, \theta)$ becomes:
\begin{eqnarray}
\label {TFT 2cyl} 
{ik}\frac {\partial E(z, r, \theta)}{\partial z} + 
\frac {1}{r} \frac {\partial}{\partial r} 
r \frac {\partial {E(z, r, \theta)}}{\partial r} +
& \nonumber \\ 
\frac {1}{r^2} \frac {\partial^2 {E(z, r, \theta)}}
{{\partial {\:}\theta}^2}+
 k^2 n (z, r, \theta ) {E(z, r, \theta)}= 0 {\:}{\:} ,{\:}{\:}{\:}{\:}
\end{eqnarray}

in free space ($n (z, r, \theta))=1$) for an arbitrary plane $z$ 
the exact solution for an $axially$ $symmetric$ beams 
reads as follows:
\begin{eqnarray}
\label {TFT 3cyl}
E(z, r)= \frac {i e^{ikz}}{\lambda z} 
\int^{\infty}_{0} E(z=0,{r}'){r}' d {r}' 
& \nonumber \\ 
\int^{2\pi}_0 \exp {\lbrace - \frac {ik}{z}{r}'\cdot r 
\cos (\theta-\theta')\rbrace }
d{\theta'}{\:},{\:}{\:}{\:}{\:}
\end{eqnarray}
where using the properties of Bessel functions 
\cite {Vladimirov:1971} we have:
\begin{equation}
\label {TFT 3cyl22}
E(z, r)= \frac {2\pi i e^{ikz}}{\lambda z} 
\int^{\infty}_{0} E(z=0,{r}'){r}' J_0 (\frac {k r' r}{z}) d {r}'
{\:}.{\:}{\:}{\:}
\end{equation}

Consider axially symmetric diaphragm $D(\vec r)=D(r)$ placed coaxially with 
axially symmetric beam $E(z,{r})$. Then beam propagation functional 
for axially symmetric geometry $T_{ax}$ is:
 \begin{eqnarray}
\label {TFT from solutions3axial } 
T_{ax} [\psi ]  =  \frac {\epsilon_0 c{\:} 2\pi k^2}{4 \pi^2 z^2} 
\int_{0}^{\infty} \int_{0}^{\infty}
A({r}')A({r}'')D_{iHT}({r}'-{r}'')
& \nonumber \\ 
\exp \lbrace i [\psi({r}')-\psi( {r}'')] 
\rbrace 
d{{r}''}d{{r}'}{\:} , {\:}{\:}{\:}
\end{eqnarray}
where kernel $D_{iHT}({r}'-{r}'')$ is $inverse$ 
Hankel transform 
of $D(r)$ or far field diaphragm as it viewed from plane $z=0$. 
In axially symmetric case the functional 
$T_{ax}[\psi ]$ is positively defined as well and we have 
the real part of $T_{ax}[\psi ]$ 
or a power flux through $D( r) $: 
 \begin{eqnarray}
\label {TFT from solutions cyl} 
Re \lbrace {T_{ax} [\psi ] \rbrace} = \frac {\epsilon_0 c{\:}2\pi  k^2}{4 \pi^2 z^2}    
\int_{0}^{\infty} \int_{0}^{\infty}
A({r}')A({r}'')D_{iHT}({r}'-{r}'')
& \nonumber \\ 
\cos \lbrace  [\psi({r}')-\psi( {r}'')] 
\rbrace 
r'' r' d{{r}''}d{{r}'}{\:}. {\:}{\:} {\:}{\:}{\:}{\:}
\end{eqnarray}
In the same way as for Cartesian coordinates the 
imaginary part of $T_{ax}[\psi ]$ is exactly zero :

 \begin{eqnarray}
\label {TFT from solutions cyl} 
Im \lbrace {T_{ax} [\psi ] \rbrace} =  0{\:}. {\:} {\:} {\:} 
\end{eqnarray}
This seemingly trivial statement on exact zero for imaginary part of the functionals 
$T [\psi ]$ and  $T_{ax} [\psi ]$ is a key point for 
evaluation of the second variational derivative.

\section {Stationary points of functional}
 
The standard procedure for searching the stationary points of 
functional is in calculation of variational derivatives, 
finding stationary solutions under condition of the first variational 
derivative equals zero and studying the sign of second variational 
derivative to decide whether extrema are 
maximas or minimas \cite {Vladimirov:1971}. The variations to the first order $\delta \psi$ : 

\begin{equation}
\label {first derivative TFT} 
\frac {\delta T [\psi]}{\delta \psi}=0 {\:}{\:}   ,{\:}{\:}
\frac {\delta^2 T [\psi]}
{{\delta \psi}^2}= {\:}{\:}? {\:}{\:} \lessgtr 0 ?
\end{equation}
In our case let us take infinitesimal trial deviation function 
$\delta \psi (\vec {r}')$ which perturbes the 
extremal function to be found 
$\psi (\vec {r}')$. Next expansion of $T [\psi + \delta \psi]$ in  
Taylor series is as follows: 

\begin{eqnarray}
\label {expansion TFT from solutions 44 } 
T [\psi + \delta \psi]  = \frac {\epsilon_0 c{\:} k^2}{4 \pi^2 z^2} 
\int_{-\infty}^{\infty} \int_{-\infty}^{\infty}
A(\vec {r}')A(\vec {r}'')D_{iFT}(\vec {r}'-\vec {r}'')
& \nonumber \\ 
\cos\lbrace [\psi(\vec {r}')-\psi(\vec {r}'')+\delta \psi (\vec {r}')-
\delta \psi (\vec {r}'')] 
\rbrace 
d^2{\vec {r}''}d^2{\vec {r}'}{\:}{\:} , {\:}{\:}{\:}{\:}{\:}{\:}
\end{eqnarray}
  
The expansion to the perturbations of the second order has 
the following three terms:

\begin{eqnarray}
\label {expansion TFT from solutions5} 
T [\psi + \delta \psi]  = \frac {\epsilon_0 c{\:} k^2}{4 \pi^2 z^2}  
\int_{-\infty}^{\infty} \int_{-\infty}^{\infty}
A(\vec {r}')A(\vec {r}'')D_{iFT}(\vec {r}'-\vec {r}'')
& \nonumber \\ 
\lbrace \cos [\psi(\vec {r}')-\psi(\vec {r}'')] -
\sin  [\psi(\vec {r}')-\psi(\vec {r}'')] \cdot 
[\delta \psi (\vec {r}')- \delta \psi (\vec {r}'')] 
& \nonumber \\ 
-\frac {1}{2!} \cos[ \psi (\vec {r}')-  \psi (\vec {r}'')] \cdot 
[\delta \psi (\vec {r}')- \delta \psi (\vec {r}'')]^2\rbrace 
d^2{\vec {r}''}d^2{\vec {r}'}+...{\:}.{\:}{\:}{\:}{\:}
\end{eqnarray}

The first term is a power flux through aperture without phase 
perturbations $\delta \psi (\vec {r}'),\delta \psi (\vec {r}')$. 
The second term contains the first variational derivative 
of the $T [\psi]$ in Cartesian coordinates: 

\begin{eqnarray}
\label {first derivative TFT full} 
\frac {\delta T [\psi]}{\delta \psi}= 
- \frac {\epsilon_0 c{\:} k^2}{4 \pi^2 z^2} \int_{-\infty}^{\infty} \int_{-\infty}^{\infty}
A(\vec {r}')A(\vec {r}'')D_{iFT}(\vec {r}'-\vec {r}'')
& \nonumber \\ 
\sin  [\psi(\vec {r}')-\psi(\vec {r}'')] 
d^2{\vec {r}''}d^2{\vec {r}'},{\:}{\:}{\:}{\:}
\end{eqnarray}

and in cylindrical coordinates for $T_{ax}[\psi]$ :

\begin{eqnarray}
\label {first derivative TFT full_cylind} 
\frac {\delta T_{ax} [\psi]}{\delta \psi}= -\frac {\epsilon_0 c{\:}2\pi  k^2}{4 \pi^2 z^2} 
\int_{0}^{\infty} \int_{0}^{\infty}
A({r}')A({r}'')D_{iHT}({r}'- {r}'')
& \nonumber \\ 
\sin  [\psi({r}')-\psi( {r}'')] r'' r' d{{r}''}d{{r}'},{\:}{\:}{\:}{\:}
\end{eqnarray}

This nonlinear integral equation 
(\ref{first derivative TFT full}) has 
a simple $continuum$ set of solutions $\psi (\vec {r})=const$ 
because for these uniform phase distributions 
the phase-modulation term $\sin  [\psi(\vec {r}')-\psi(\vec {r}'')] $ 
is exactly zero and the whole integral is zero as well. Thus it is 
clear that solutions $\psi (\vec {r})=const$ correspond to extrema 
of the beam propagation functional. 

In order to detect the kind of extrema 
the second variational derivative is to be analyzed. 
When second derivative is positive (\ref{first derivative TFT}) 
the functional is $concave$ and 
this means it has the $local$ minimum on this set of functions.  
Vice versa the $convex$ functional 
with the $negative$ second variational derivative 
has at least a $local$ maximum on the given solution $\psi (\vec {r})$. 

The second order variations are inside the third term 
of expansion in cartesian coordinates(\ref {expansion TFT from solutions5}): 

\begin{eqnarray} 
\label {second derivative TFT full} 
 {\delta^2 T [\psi]}= - 
\frac {1}{2!} \frac {\epsilon_0 c{\:} k^2}{4 \pi^2 z^2} 
 \int_{-\infty}^{\infty} \int_{-\infty}^{\infty}
A(\vec {r'})A(\vec {r''})D_{iFT}(\vec {r'}-\vec {r''})
& \nonumber \\ 
\cos [\psi(\vec {r'})-\psi(\vec {r''})] \times 
[\delta \psi (\vec {r'})- \delta \psi (\vec {r''})]^2 \cdot  
d^2{\vec {r''}}d^2{\vec {r'}} ,{\:}{\:}{\:}{\:}
\end{eqnarray}

and in polar coordinates for $T_{ax}$: 

\begin{eqnarray}
\label {second derivative TFT full_cylind} 
 {\delta^2 T_{ax} [\psi]}= - 
\frac {2\pi }{2!} \frac {\epsilon_0 c{\:} k^2}{4 \pi^2 z^2}  
\int_{0}^{\infty} \int_{0}^{\infty}
A({r'})A({r''})D_{iHT}({r'}- {r''})
& \nonumber \\ 
\cos  [\psi({r'})-\psi( {r''})] \times 
[\delta \psi ( {r'})- \delta \psi ({r''})]^2 
r'' r' d{{r''}}d{{r'}},{\:}{\:}{\:}{\:}
\end{eqnarray}

The integrals (\ref{second derivative TFT full}) and 
(\ref{second derivative TFT full_cylind})
are positive because they contain the positively defined 
squares $[\delta \psi (\vec {r'})- \delta \psi (\vec {r''})]^2$ 
and $[\delta \psi ({r'})- \delta \psi ({r''})]^2$ 
of an infinitesimally small $real$ trial functions 
$\delta \psi (\vec {r'}), \delta \psi (\vec {r''})$ inserted inside 
$positive$ integral of the total power flux 
(\ref {TFT from solutions 4}) through aperture $D(\vec r)$.

The second variational derivative is given by a third term 
of expansion in cartesian coordinates(\ref {expansion TFT from solutions5}): 

\begin{eqnarray}
\label {second derivative TFT full22} 
\frac {\delta^2 T [\psi]} {{\delta \psi}^2}= 
-\frac {1}{2!} \frac {\epsilon_0 c{\:} k^2}{4 \pi^2 z^2} 
 \int_{-\infty}^{\infty} \int_{-\infty}^{\infty}
A(\vec {r}')A(\vec {r}'')\cdot
& \nonumber \\
D_{iFT}(\vec {r}'-\vec {r}'')\cdot \cos [\psi(\vec {r}')-\psi(\vec {r}'')]\cdot  
d^2{\vec {r}''}d^2{\vec {r}'} ,{\:}{\:}{\:}{\:}
\end{eqnarray}

and in polar coordinates for $T_{ax}$ (\ref {expansion TFT from solutions5}): 

\begin{eqnarray}
\label {second derivative TFT full_cylind22} 
\frac {\delta^2 T_{ax} [\psi]}{\delta \psi^2}= - 
\frac {2\pi }{2!} \frac {\epsilon_0 c{\:} k^2}{4 \pi^2 z^2}  
\int_{0}^{\infty} \int_{0}^{\infty}
A({r}')A({r}'')\cdot
& \nonumber \\
D_{iHT}({r}'- {r}'')\cdot \cos  [\psi({r}')-\psi( {r}'')] \cdot 
r'' r' d{{r}''}d{{r}'}.{\:}{\:}{\:}{\:}
\end{eqnarray}

The integrals for the second variational derivative (\ref{second derivative TFT full22}) 
are positively defined for $\psi ({\vec r }')=const$
because 
integral of the total power flux 
(\ref {TFT from solutions 4}) through aperture $D(\vec r)$ is $positive$ . 
Consequently because of multiplication by negative $-\frac {1}{2!} $ it is 
clear that second variational derivative is negative for all 
infinitesimal phase perturbations  
and beam propagation functional $T [\psi ] $ is $convex$ in the vicinity 
of exact solutions $\psi (\vec {r})=const$. 
This means that homogeneous 
phase distributions $\psi (\vec {r})=const$ in near field 
provide $at$ $least$ a $local$ maximum of the optical 
flux transmitted 
through aperture $D(\vec r)$ located in far field. 

\begin{figure} 
\center{ \includegraphics[width=9 cm]{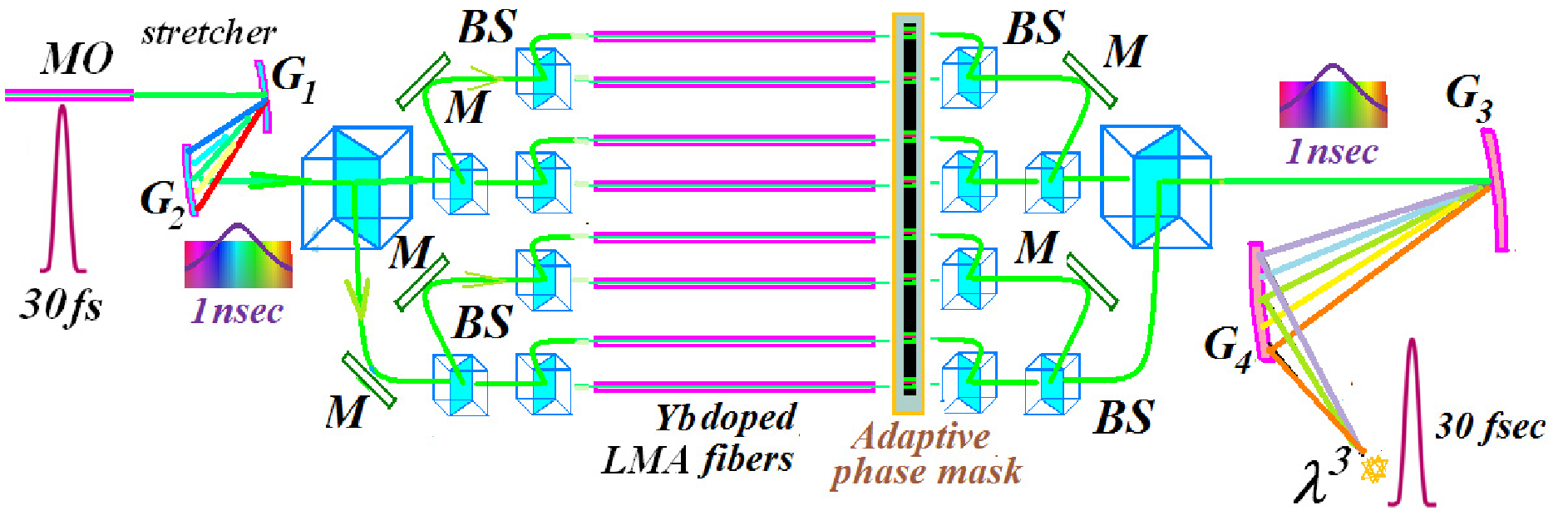}}
\caption{ (Color online) Phase-locked LMA (large mode area) 
chirped pulse laser network \cite {Mourou:2013}. $MO$- master 
oscillator emitting 30-$fs$ transform-limited pulses, the $stretcher$ 
elongates pulses to 1 $ns$, 
and $compressor$ compresses amplified chirped pulses to initial 30-$fs$,
each with a pair of diffraction gratings $G_1, G_2$ 
and $G_3, G_4$. Binary trees of beamsplitters $BS$ and mirrors $M$
provide beam multiplexing and beam combination. Array of large mode area $LMA$ 
fiber amplifiers and wavefront control $adaptive$ $phase$ $mask$ with 
$\lambda/(10 \div 100)$ accuracy form the diffraction-limited 
focal region with $\lambda^3$ volume.} 
\label{fig.3}
\end{figure}

In the same way as in the many other mini-max problems the maximum of optical 
flux through aperture $D(\vec r)$ ia far field means the $minimum$ 
of the optical flux beyond the aperture $T_{bey} [\psi (\vec r,z) ]$. 
In this case the second variational 
derivative of functional : 
\begin{eqnarray}
\label {TFT_beyond} 
T_{bey} [\psi (\vec r,z) ]  =   \int_{-\infty}^{\infty} (1-D(\vec r)) I (\vec r,z)  
d^2{\vec r}{\:}{\:};
& \nonumber \\ 
\frac {\delta T_{bey} [\psi (\vec r,z)]}{\delta \psi}=0    ,{\:}{\:}
\frac {\delta^2 T [\psi]}
{{\delta \psi}^2} {\:} >{\:} 0  {\:}{\:} 
{\:}{\:}{\:}{\:}
\end{eqnarray}

is exactly positive in the vicinity of extremum.
 
\section{Conclusion}

The exact solution of variational problem for laser beam concentration in a 
focal plane located on beam axis had been obtained for an arbitrary realistic 
initial conditions in the form of fundamental Gaussian modes, 
spatial solitons \cite{Okulov:2020},\cite{Okulov:1988} and 
phase-locked laser arrays \cite{Okulov:1990,Mourou:2013}. 
The optimal phase distribution in the near field had been 
found for $fixed$ light amplitude distribution. The solutions 
(\ref{first derivative TFT full}-\ref{second derivative TFT full_cylind})
are relevant to the numerous 
applications of Shack-Hartmann technique including 
astronomical mirrors and phase-locked fiber lasers\cite{Mourou:2013} 
traditionally 
used for adjustment and compensation of the tiny wavefront 
distortions \cite {Basov:1980,Okulov:1983} via lenslet 
arrays sensing and high-performance computing \cite {Mourou:2013}. 

For the first sight the result on the maximum of the 
optical flux $T[\psi (\vec r,z)\rightarrow \infty  ]$ 
through an aperture placed in the far field $D(\vec r)$ (fig.2)
looks quite obvious because there exist the common wisdom 
of a well-known property of Fourier series and Fourier transforms that 
homogeneous phase distribution across spectrum means the minimal spectral  
width or else the transform limited pulses have a minimal duration in 
the absence of the phase-modulations. Indeed this common sense is 
well known to the most of researchers as a result of their own experience 
with calculations and measurements Fourier spectra rather than within 
rigorous proof. Hopefully the above approach based on $exact$ formulas from scalar 
diffraction theory $\cite{Born_Wolf:1972}$ provides the additional  
side-view upon signal and spectra formation tightly linked wirh quantum-classical 
correspondence and particle-wave duality issues.

Compared to the previously formulated variational approaches relevant 
to soliton formation \cite{Malomed:2002}, wave propagation in 
rectangular waveguides \cite{Haus:1982} and in a bulk nonlinear 
medium \cite{Rubinstein:2004} the results presented above 
give the necessary and sufficient conditions for laser power maximization 
in a far field and in a focal spot of objective.

\section{Disclosures}
Author declare no conflicts of interest.

\section{Data availability statement}
The data that support the findings of this study are
available from the author upon reasonable request.

\end{document}